\documentclass[aps,prl,twocolumn,showpacs,groupedaddress,amsmath,amssymb]{revtex4}
\usepackage{graphicx}
\usepackage{dcolumn}
\usepackage{bm}
\usepackage{psfig,graphics}
\usepackage{latexsym}
\usepackage{amsmath}
\usepackage{amssymb}
\usepackage{enumerate}
\usepackage[latin1]{inputenc}
\usepackage{bbm}

\newcommand{\bra}[1]{\ensuremath{\langle#1|}}
\newcommand{\Ket}[1]{\ensuremath{|#1\rangle}}
\newcommand{\ket}[1]{\ensuremath{|#1\rangle}}
\newcommand{\BraKet}[2]{\ensuremath{\langle #1|#2\rangle}}

\newcommand{\KetBra}[1]{\ensuremath{| #1 \rangle \langle #1 |}}
\newcommand{\ketbra}[1]{\ensuremath{| #1 \rangle \langle #1 |}}

\newcommand{\Eins}{\ensuremath{\mathbbm{1}}}
\newcommand{\eins}{\ensuremath{\mathbbm{1}}}

\newcommand{\WW}{\ensuremath{\mathcal{W}}}

\newcommand{\kommentar}[1]{}

\begin{document}


\title{Experimental Detection of  Multipartite Entanglement using Witness Operators}
\author{Mohamed Bourennane$^{1,2}$, Manfred Eibl$^{1,2}$, Christian
Kurtsiefer$^{2}$, Sascha Gaertner$^{1,2}$, Harald
Weinfurter$^{1,2}$, Otfried G\"uhne$^{3}$, Philipp Hyllus$^{3}$,
Dagmar  Bru\ss$^{3}$, Maciej Lewenstein$^{3}$, and Anna
Sanpera$^{3}$} \affiliation{$^{1}$Max-Planck-Institut f\"{u}r
Quantenoptik, D-85748 Garching, Germany} \affiliation{$^{2}$Sektion
Physik, Ludwig-Maximilians-Universit\"{a}t, D-80797 M\"{u}nchen,
Germany} \affiliation{$^{3}$Institut f{\"u}r Theoretische Physik,
Universit{\"a}t Hannover, D-30167 Hannover, Germany}
\date{\today}
\begin{abstract}
We present the experimental detection of {\it genuine}
multipartite entanglement using entanglement witness operators. To
this aim we introduce a canonical way of constructing and
decomposing  witness operators so that they can be directly
implemented  with present technology. We apply this method to
three- and four-qubit entangled states of polarized photons,
giving experimental evidence that the considered states contain
true multipartite entanglement.
\end{abstract}
\pacs{03.67.Mn,
 03.65.Ud,
03.67.-a.}

\maketitle
\date{\today}
\newpage


Entanglement is one of the most puzzling features of quantum
theory and of great importance for the new field of  quantum
information theory. The determination of whether a given state is
entangled or not is one of the most challenging open problems of
this field. For the experimental detection of entanglement Bell
inequalities~\cite{B64} are widely used. However, even for
two-qubit systems there exist entangled states which do not
violate any Bell inequality~\cite{werner}.
The tool of choice in this case is the Peres-Horodecki
criterion~\cite{peres,horos} as it gives a simple sufficient and
necessary condition for entanglement.
Yet, the situation is much more complicated for higher dimensional
and multipartite systems, where simple necessary and sufficient
conditions are not known \cite{giedke}.

In the analysis of multipartite systems, it is important to
distinguish between {\em genuine} multipartite entanglement and {\em
biseparable} (triseparable, etc.) entanglement. Genuine multipartite
entangled pure states cannot be created without participation of all
parties. Conversely, for pure biseparable states of $n$ parties a
group of $m<n$ parties can be found which are entangled among each
other, but not with any member of the other group of $n-m$
parties~\cite{mixedstates}. Distinction and detection of genuine
multipartite entanglement poses an important challenge in quantum
information science. Bell inequalities are not suited to this aim in
general. Multiseparable  and biseparable states violate known Bell
inequalities less than $n$-partite Greenberger-Horne-Zeilinger (GHZ)
states.
However, for $n>3$ there exist even pure $n$-partite entangled
states with a lower violation than biseparable
states~\cite{collins}. Only recently, significant progress in
classifying multipartite entanglement has been achieved using
entanglement witnesses~\cite{horos,witnesses}. These observables
can {\em always} be used to detect various forms of multipartite
entanglement, when some {\it a priori} knowledge about the states
under investigation is provided \cite{detec2}; they are in this
sense more powerful than Bell inequalities.

A witness of genuine $n$-partite entanglement is an observable
which has a positive expectation value on states with $n-1$
partite entanglement and a negative expectation value on some
$n$-partite entangled states. The latter states and their
entanglement, respectively, are said to be detected by $\WW$.
Witnesses provide sufficient criteria for
entanglement and for distinguishing the various classes of
genuine multipartite entangled states.

The goal of this Letter is twofold. First, we introduce a general
scheme for the construction of multipartite witness operators and
their decomposition into locally measurable observables. In this
way, we demonstrate how witness operators can be implemented
experimentally in a straightforward way by using local projective
measurements, even for multipartite systems \cite{fdm}. Then, we
apply this scheme to certain states and perform the experimental
detection of their multipartite entanglement, which could not be
revealed by known Bell inequalities. In particular, we use this
method for the characterization of the three-qubit W
state~\cite{duer}, and the four-qubit state
$\ket{\Psi^{(4)}}$~\cite{weinzuk}.

A witness operator that detects  genuine multipartite entanglement
of a pure state $\ket{\psi}$
(and of states that are close to $\ket{\psi}$, e.g. in the presence of noise)
is given by
\begin{equation}
\WW=\alpha \Eins - \KetBra{\psi} \ ,
\label{witnessdef}
\end{equation}
where $\Eins$ is the identity operator,
\begin{equation}
\alpha = \max_{\ket{\phi} \in {B}}|\BraKet{\phi}{\psi}|^2 \ ,
\label{alpha}
\end{equation}
and $B$ denotes the set of biseparable states. This construction
guarantees that ${\rm Tr}(\WW \rho_{B}) \geq 0$ for all biseparable
states $\rho_{B},$ and that ${\rm Tr}(\WW \ketbra{\psi}) < 0.$ Thus,
a negative expectation value of the observable $\WW$ clearly
signifies that the state $\ket{\psi}$ carries multipartite
entanglement. Determining $\alpha$ in Eq. (\ref{alpha}) is a
difficult task, when the maximization of the overlap with {\em any}
biseparable state is performed explicitly. However, a simple method
based on the Schmidt decomposition of bipartite partitions was
found; details are described in the Appendix.

For the experimental implementation of the witness
(\ref{witnessdef}), it is necessary to decompose it into a number
of {\em local} von Neumann (or projective) measurements
\cite{detec}
\begin{equation}
\WW=\sum_{k=1}^K M_k \label{optdec}\ ,
\end{equation}
where
\begin{equation}
M_k = \sum_{l_1,...,l_n}d_{l_1,...,l_n}^{(k)}
\ketbra{a^{(k,1)}_{l_1}}\otimes... \otimes\ketbra{a^{(k,n)}_{l_n}}
\label{lvnmdef}.
\end{equation}
Here $n$ is the number of parties, $\ket{a^{(k,m)}_{l_m}}$ are
orthogonal vectors for a fixed $(k,m)$, and $d^{(k)}_{l_1...l_n}$
are real weighting coefficients. An observable $M_k$ can be
measured with {\em one} setting of the measuring devices of the
parties. We call the local decomposition (\ref{optdec})
``optimal'' when $K$ is minimal.

For demonstrating the power of multipartite witnesses we choose
states with non-maximal multipartite entanglement. The first one
is the three-qubit W state \cite{duer, kiesel}
\begin{equation}
\ket{W} = \sqrt{\frac{1}{3}}(\ket{001} + \ket{010} + \ket{100})
\label{W} \ .
\end{equation}
It has been shown that this state is not equivalent to the GHZ
state under stochastic local operations and classical
communication \cite{duer} and possesses different entanglement
properties than the GHZ state. The second state is the four-qubit
state $\ket{\Psi^{(4)}}$, which is a superposition of a four-qubit
GHZ state and the tensor product of two maximally entangled
two-qubit states. It is given by \cite{weinzuk,Eibl}
\begin{eqnarray}
\ket{\Psi^{(4)}} &=& \frac{1}{\sqrt{3}}\Big(\ket{0011}+\ket{1100}
\nonumber \\ &-&
\frac{1}{2}(\ket{0110}+\ket{1001}+\ket{0101}+\ket{1010}) \Big)
\label{psi4}.
\end{eqnarray}

For the entanglement detection of the three-qubit W state
(\ref{W}) we have considered two witnesses $\mathcal{W}_{W}^{(1)}$
and $\mathcal{W}_{W}^{(2)}$  \cite{acin}. The first witness is
constructed according Eq. (\ref{witnessdef}), resulting in
\begin{eqnarray}
\mathcal{W}_{W}^{(1)}& = &\frac{2}{3}\eins - \ketbra{W} \nonumber
\\ &=& \frac{1}{24} \Bigl[ 17 \cdot  \Eins^{\otimes 3} + 7 \cdot
\sigma_z^{\otimes 3} + 3 \cdot \bigl( \sigma_z  \Eins  \Eins +
  \Eins  \sigma_z  \Eins
\nonumber \\ &+& \Eins \Eins \sigma_z \bigr) + 5 \cdot \bigl(
\sigma_z \sigma_z  \Eins +
      \sigma_z  \Eins  \sigma_z+
      \Eins \sigma_z  \sigma_z \bigr)
\nonumber\\ &-&
  (\Eins+\sigma_z+\sigma_x)^{\otimes 3}
- (\Eins+\sigma_z-\sigma_x)^{\otimes 3} \nonumber\\ &-&
 (\Eins+\sigma_z+\sigma_y)^{\otimes 3}
- (\Eins+\sigma_z-\sigma_y)^{\otimes 3} \Bigr] \:. \label{Ww1}
\end{eqnarray}
Its expectation value is positive on biseparable and fully separable
states. It thus detects all states belonging to the two classes of
states with genuine tripartite entanglement, the W class and the GHZ
class, but without distinguishing between them. The factor $2/3$
corresponds to the maximal squared overlap between the W state and
biseparable states. From this we also see that a mixture of
$\ket{W}$ and white noise, $\rho = p\ketbra{W}+(1-p)\eins/8,$
exhibits tripartite entanglement for a noise contribution of up to
$p>13/21$. We introduced a short notation for tensor products, {\it
i.e.,} $\eins \sigma_i \sigma_j := \eins \otimes \sigma_i \otimes
\sigma_j$, where $\Eins$ and $\sigma_i$ are the identity and Pauli
matrices. This decomposition requires five measurement settings,
namely $\sigma_z^{\otimes 3}$ and
$((\sigma_z\pm\sigma_i)/\sqrt{2})^{\otimes 3}; i=x,y$, see also
Fig.~\ref{ww}(below). Finding such a decomposition and proving its
optimality is technically demanding (for details see
~\cite{detec2}).

A witness that detects genuine tripartite entanglement and, with the
same set of local measurements, allows to distinguish between the W
and GHZ states, is given by~ \cite{acin}
\begin{eqnarray}
\mathcal{W}_{W}^{(2)} &=& \frac{1}{2}\eins - \ketbra{\overline{GHZ}}
\nonumber \\ &=& \frac{1}{16} \Bigl[ 6 \cdot \eins^{\otimes 3} + 4
\cdot \sigma_z^{\otimes 3} -  2 \cdot \bigl( \sigma_y  \sigma_y
\eins +
   \sigma_y \eins  \sigma_y
\nonumber \\ &+& \eins \sigma_y  \sigma_y \bigr) - (\sigma_z
+\sigma_x)^{\otimes 3}- (\sigma_z -\sigma_x)^{\otimes 3} \Bigr]
\label{Ww2} ,
\end{eqnarray}
where $\ket{\overline{GHZ}} = (\ket{\bar{0}\bar{0}\bar{0}}
+\ket{\bar{1}\bar{1}\bar{1}})/\sqrt{2} =
(\ket{000}+\ket{001}+\ket{010}+\ket{100})$ with $\ket{\bar{0}} =
(\ket{0}+i\ket{1})/\sqrt{2}$ and
$\ket{\bar{1}}=-(\ket{0}-i\ket{1})/\sqrt{2}$. This witness is
constructed slightly differently from above, namely here $1/2$ is
the maximal squared overlap between $\ket{\overline{GHZ}}$ and any
biseparable state. Furthermore, since the maximum overlap between
$\ket{\overline{GHZ}}$ and any W state is $3/4$ ~\cite{acin}, the
operator $\WW_{GHZ}=3/4\cdot\Eins-\KetBra{\overline{GHZ}}$ is a
GHZ witness, {\it i.e.,} it  has a negative expectation value for
GHZ states, but is positive for states belonging to the class of W
states. Therefore we can prove with the witness (\ref{Ww2}) that a
state $\rho$ is fully tripartite entangled if
${\rm Tr}(\mathcal{W}_{W}^{(2)}\rho) < 0$. If
${\rm Tr}(\mathcal{W}_{W}^{(2)}\rho) < -1/4$ then the state $\rho$ does
not belong to the W state class.
Theoretically, one expects $ {\rm Tr}(\mathcal{W}_{W}^{(1)}\ketbra{W}) =
-{1}/{3}$ and $ {\rm Tr}(\mathcal{W}_{W}^{(2)}\ketbra{W}) = -1/4.$

For the four-qubit entangled state (\ref{psi4}), we use the
witness
\begin{eqnarray}
\WW_{\Psi^{(4)}} & = & \frac{3}{4}\eins -\ketbra{\Psi^{(4)}}
\\
&=& \frac{1}{48} \Bigl( 3 \cdot (\sigma_x \sigma_x  \sigma_y
\sigma_y +  \sigma_y  \sigma_y \sigma_x \sigma_x +  \sigma_x
\sigma_x \sigma_z \sigma_z)
\nonumber \\
&+& 3 \cdot (\sigma_z \sigma_z  \sigma_x  \sigma_x +  \sigma_y
\sigma_y  \sigma_z  \sigma_z +  \sigma_z \sigma_z  \sigma_y
\sigma_y)
\nonumber \\
&-&
  (\sigma_x + \sigma_y)^{\otimes 4}
- (\sigma_x -\sigma_y)^{\otimes 4} - (\sigma_x +
\sigma_z)^{\otimes 4}
\nonumber \\
&-&
  (\sigma_x -\sigma_z)^{\otimes4}
- (\sigma_x +\sigma_z)^{\otimes 4} - (\sigma_y
-\sigma_z)^{\otimes4}
\nonumber \\
&+& 33 \cdot \eins^{\otimes4} - \sum_{i=x,y,z} \bigl[ \sigma_i
\sigma_i  \eins  \eins +\eins  \eins  \sigma_i  \sigma_i
-\sigma_i^{\otimes4}
\nonumber \\
&-& 2 \cdot (\sigma_i   \eins  \sigma_i  \eins  +
 \sigma_i  \eins  \eins  \sigma_i +
 \eins  \sigma_i  \sigma_i  \eins +
 \eins  \sigma_i  \eins \sigma_i)
\bigr]\Bigr).\nonumber
\end{eqnarray}
This decomposition requires 15 measurement settings. The witness
$\WW_{\Psi^{(4)}}$ has a positive expectation value on all
triseparable, biseparable, and fully separable states. Here, the
theoretical expectation value is given by
${\rm Tr}(\mathcal{W}_{\Psi^{(4)}}\rho_{\Psi^{(4)}})= -1/4.$
All of the above witnesses can detect the desired states mixed
with not too large
noise.

Let us now proceed with the experimental demonstration. We have
chosen  multi photon states as the test bench of the entanglement
witness. For these experiments the qubits are represented by the
polarization of the photons, with $"0" \equiv$ horizontal (H) and
$"1" \equiv$ vertical (V) linear polarization. The process of
spontaneous parametric down-conversion (SPDC) is used to generate
a polarization-entangled four photon state in the arms $a_0$ and
$b_0$ \cite{DC95,weinzuk} (Fig. 1).
For enabling four observers in the arms $a,b,c$ and $d$ to analyze
the $\Psi^{(4)}$ state it suffices to distribute the four photons
using semitransparent beam splitters (BS) (Fig. 1(a)). To transform
the initial state into the $W$ state we employ two photon
interference at a BS when distributing the photons into arms $a,b$
and $c$ (Fig. 1(b)) \cite{Wexp}. Provided each of the three (four)
observers receive one photon they obtain the three photon state
$\ket{W}$, or the four photon state $\ket{\Psi^{(4)}}$ respectively.
The general priciple and the experimental techniques to observe
multi photon entangled states are described in detail in
\cite{Wstateexp,Eibl}; let us here focus on detecting their
entanglement.

\begin {figure}
\includegraphics[width=8.5cm]{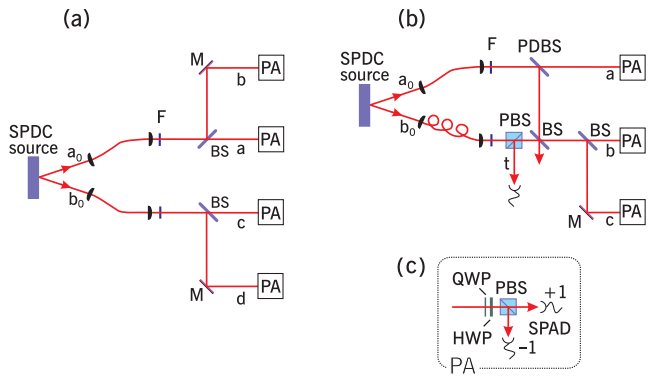}
\caption{Experimental setups to demonstrate (a) four-photon
entanglement of the $\Psi^{(4)}$ state and (b) three-photon
entanglement of the W state; (c) polarization analysis (PA)
setup.} \label{fig:Wsetup}
\end{figure}

For implementing the witness observable polarization analyzers (PA)
are used. A quarter- (QWP) and a half-wave-plate (HWP) together with
a  (PBS) allow to set and to analyze any arbitrary polarization
direction of each of the photons. As the computational basis of the
qubit $"0"/"1"$ and thus the spin observable $\sigma_z$ corresponds
to a measurement of the H/V linear polarization, $\sigma_x$
($\sigma_y$) corresponds to the analysis of $\pm45^\circ$ linear
polarization (left/right circular polarization). Registration of a
photon in one of the two detectors of a PA signals the observation
of the corresponding eigenstate of the spin operator. Every possible
observable of the type
$\ketbra{a^{(k,1)}_{l_1}}\otimes\ldots\otimes\ketbra{a^{(k,n)}_{l_n}}$
(4) corresponds to one of the $2^n$ possible detection events where
each of the $n$ observers registers one photon, either with
eigenvalue $l_i=+1$ or $-1$. From the probability of these multi
photon detections, $P(a^{(k,1)},\ldots, a^{(k,n)})_{l_{1},\ldots,
l_{n}}$, we then can compute the various terms ${\rm
Tr}(M_{k}\rho)=\sum_{l_1,\ldots, l_n=\pm 1} d^{(k)}_{l_{1},\ldots,
l_{n}} P(a^{(k,1)},\ldots, a^{(k,n)})_{l_{1},\ldots, l_{n}}$ of the
entanglement witness expectation value.


The multi photon detection probabilities for the three-qubit state
$\ket{W}$ are shown in Fig.~\ref{ww}. From the  experimental
results we obtain
\begin{eqnarray}
{\rm Tr}(\mathcal{W}_{W}^{(1)}\rho_{W})_{exp} &=&  -0.197 \pm 0.018 \ , \\
{\rm Tr}(\mathcal{W}_{W}^{(2)}\rho_{W})_{exp} &=&  -0.139 \pm 0.030 \ .
\end{eqnarray}
This clearly proves with high statistical significance that the
observed state is truly tripartite entangled. We want to emphasize
that the evaluation of a three-photon Bell inequality failed to
signify tripartite entanglement for the same experimental settings
and noise~\cite{Wstateexp}, indicating the superiority of the
witnesses.

\begin {figure}
\includegraphics[width= 8.5cm]{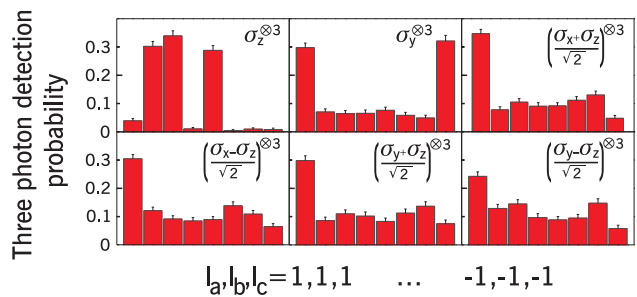}
\caption{Three photon detection probabilities for six settings of
the polarization analyzers as required for the detection of
three-photon entanglement using the witness operators
$\mathcal{W}_{W}^{(1)}$ and $\mathcal{W}_{W}^{(2)}$.}
 \label{ww}
\end{figure}

\begin {figure}
\includegraphics[width = 8.5cm]{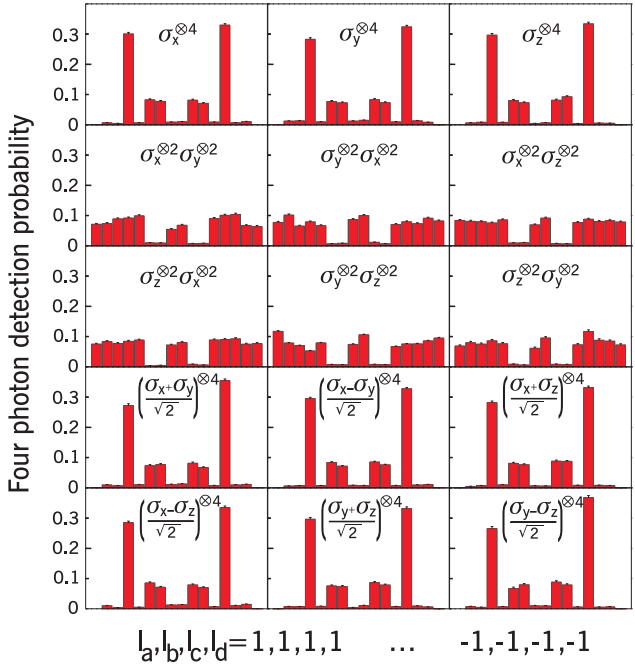}
\caption{Four photon detection probabilities for fifteen settings
of polarization analyzers as required for the detection of
four-photon entanglement using the witness operators
$\mathcal{W}_{\Psi^{(4)}}$ .}
 \label{wzfig}
\end{figure}

For the detection of the genuine fourpartite entanglement of the
state $\ket{\Psi^{(4)}}$, 15 different analyzer settings are
required. This is comparable with the 16 settings required for
evaluating a four-photon Bell inequality~\cite{weinzuk}. However,
for $\ket{\Psi^{(4)}}$ all known Bell inequalities give a
violation which is lower than the maximum one of biseparable
states, therefore only tripartite entanglement could be shown by
those violations. Thus, it is only the witness
$\mathcal{W}_{\Psi^{(4)}}$ which can prove the fourpartite
entanglement of $\ket{\Psi^{(4)}}$. The observed fourfold
detection probabilities (Fig.~\ref{wzfig}) result in an
expectation value of
\begin{eqnarray}
{\rm Tr}(\mathcal{W}_{\Psi^{(4)}}\rho_{\Psi^{(4)}})_{exp} =  -0.151 \pm
0.01 \ ,
\end{eqnarray}
which finally confirms the genuine multipartite entanglement
beyond any doubt.

In conclusion, we have developed a generic scheme to easily
construct witness operators distinguishing genuine multipartite
entangled states from biseparable states or states with even lower
multipartite entanglement. This allowed to analyze the
entanglement of three- and four-photon entangled states based on
local measurements only. Whereas the evaluation of generalized
multi-photon Bell inequalities fell short of ruling out
biseparability~\cite{Eibl,Wstateexp}, the experimentally obtained
values of the respective witnesses clearly prove the genuine
multipartite entanglement of the observed states. After solving
the problem of analyzing bipartite
entanglement~\cite{peres,horos}, we now have also a well-suited
tool at hand for the experimental analysis of genuine multipartite
entangled quantum systems.

We thank A.~Ac{\'\i}n, A.~Ekert, C.~Macchiavello, A.~Miyake, and F.~Verstraete
for discussions and acknowledge support from the Deutsche
Forschungsgemeinschaft, the Bavarian quantum information initiative
and the EU (Program RamboQ and QUPRODIS).\\

{\em Appendix. --} In this Appendix, we show how to
calculate the overlap $\alpha$ of Eq. (\ref{alpha}).
We first fix a bipartite splitting $B_1$, and
consider only states $\ket{\phi}\in B_1$ which are  product vectors with
respect to this partition. We choose an orthonormal product basis $\Ket{ij}$
for this partition, thus  $\Ket{\psi}=\sum_{ij}c_{ij}\Ket{ij}$ and
$\ket{\phi}=\ket{a}\ket{b}=\sum_{ij}a_i b_j \ket{ij}$.
The coefficient matrix is denoted by $C=(c_{ij})$
and the normalized coefficient vectors by
$\vec{a}=(a_i)$ and $\vec{b}=(b_i)$. Then
\begin{eqnarray}
\max_{\ket{\phi} \in {B_1}}|\BraKet{\phi}{\psi}|
&=&
\max_{a_i,b_j} |\sum_{ij}({a_i}^* c_{ij} {b_j}^*)|
\nonumber \\
&=& \max_{\vec{a},\vec{b}} |\bra{\vec{a}} C \ket{\vec{b}^*}| =
\max_k \{ \lambda_k (C) \} \ , \label{sup}
\end{eqnarray}
where $\lambda_k (C)$ denotes the singular values 
of $C$, {\it i.e.} the roots of the eigenvalues of $CC^\dagger$.
In other words, $\lambda_k (C)$ are the Schmidt coefficients
of $\ket{\psi}$ with respect to a fixed  bipartite splitting.
Therefore, $\alpha$ is given by the square of the Schmidt coefficient
which is maximal over {\em all} possible bipartite partitions of $\ket{\psi}$.


\begin{thebibliography}{99}
\bibitem{mixedstates} A mixed state is called biseparable if it is a
mixture of biseparable pure states; the mixture may contain terms
corresponding to different partitions. For triseparable pure
states etc. the whole state can be partitioned into three or more
such groups with consequently decreasing genuine multipartite
entanglement.
\bibitem{Wexp}
The basic idea of how to observe the $\ket{W}$ is as follows: If
one photon is registered by the trigger detector $t$ in the
reflected output of the polarizing beam splitter (PBS), and the
others in arms $a, b,$ and $c$, then two of those three photons
have horizontal polarization, with the third being vertically
polarized. For having twice the probability of observing
horizontally polarized photons compared to vertically polarized
ones, we place a polarization dependent beam splitter (PDBS) in
arm $a_0$ (with transmissions $t_H=2t_V$). The remaining two
photons are interfered at a 50-50 beam splitter (BS) and
distributed into arms $b$ and $c$ by a second BS
\cite{kiesel,Wstateexp}.
\end{thebibliography}
\end{document}